\documentstyle[12pt]{article}
\title{Massless Wess--Zumino model as \\
first quantized Siegel superparticle}
\author{A.A. Deriglazov\thanks{deriglaz@phys.tsu.tomsk.su},
\, A.V. Galajinsky\thanks{galajin@phys.tsu.tomsk.su}\\
Physics Department, Tomsk State University,\\
634050 Tomsk, Russia,}
\date{and\\
D.M. Gitman\thanks{gitman@fma.if.usp.br}\\
Instituto de Fisica, Universidade de S\~ao Paulo,\\
P.O. Box 66318, 05315-970, S\~ao Paulo, SP, Brazil.}
\begin{document}
\maketitle
\large
\begin{abstract}
It is shown that canonical quantization of the $4d$ Siegel
superparticle yields massless Wess--Zumino model as an effective field
theory. Quantum states of the superparticle are realized in terms of
the real scalar superfields which prove to be decomposed into the sum
of on-shell chiral and antichiral superfields.
\end{abstract}

\vspace{0.2cm}

\vspace{1cm}

1. Already for a long time there exists permanent interest in the study and
quantization of different classical and pseudoclassical models of
relativistic particles. In fact, such an activity began with the
works [1,2] where a pseudoclassical
model (PM) for the Dirac particle in four dimensions was constructed,
studied, and quantized. It was interesting to see how quantum mechanics
of a spinning particle, which may also be treated as a free spinor field
theory, can be reconstructed in the course of first quantization of
Grassmann classical mechanics (pseudoclassical mechanics).
It was expected that such an experience might be
useful in the intensively developed that time string theory, where
first quantization was and remains until now the main approach to quantum
description. However, it soon became clear that a PM is of independent
theoretical interest and the natural questions appeared: Is it
possible to treat any field theory (at least, a free field theory) as the
result of first quantization of a classical particle model or a PM? Does
there exist a regular method to construct such a model for a given field
theory?  Although these questions were not answered in general,
many interesting results were obtained in that direction. For
example, there were constructed PM for higher spin field theories
in 3+1 and 2+1 dimensions [3, 4]. In particular, a PM which corresponds
to the topologically massive gauge theory of spin-one particles (2+1
electrodynamics with Chern-Simons term) and a PM for the Weyl particle in
an arbitrary even dimension have been presented [5, 6].

It has long been realized that the path
integral representation for the propagator of a field theory may serve as
a heuristic method to find an action of the corresponding PM. The
effectiveness of such an approach was demonstrated for the scalar and
spinor particles in four dimensions [2, 7, 8] and then used [9] to
construct a new minimal PM for spinning particles in odd dimensions.
However, the approach mentioned can not be considered as the universal one
since the construction of a path integral itself may present a nontrivial
tack. As an example one can mention the situation with the
Weyl particle in even dimensions for which a PM was constructed but the
corresponding path integral was not yet. Generally, a path integral
representation for a field propagator keep no track of all the
constraints of the corresponding particle model. In other words, a
particle action enters into the path integral in the gauge fixed form.

It is natural to expect that a PM of a superparticle has to reproduce in
the course of first quantization some supersymmetric field theory.
However, its covariant quantization creates new
(in comparison to those of spinning particles) difficulties related to the
complicated constraints structure. Of prime interest (for motivations and a
comprehensive list of references see Ref. 10) seems to be the
Brink--Schwarz superparticle [11] which involves an infinitive ghost
tower when BRST quantized [12]. The puzzle of covariant quantization of
the superparticle can be addressed in two respects [10] (for an
alternative twistor-harmonic approach see Ref. 13 ): The problem of
quantizing the infinitely reducible second class constraints; The problem
of quantizing the infinitely reducible first class ones. It is the Siegel
superparticle [14] (later referred to as $AB$ superparticle [15]) which
allows one to examine the latter in an independent way.\footnote{In Ref.
16 further modifications ($ABC$-, $ABCD$-superparticles) have been
proposed and shown [17] to be equivalent to the Brink--Schwarz model. As
the new formulations involve first class constraints only, the second
class constraints problem intrinsic in the Brink--Schwarz theory can be
avoided (for works on quantization of the $ABCD$ superparticle see Refs.
15 and 18).}

Recently, a recipe how to supplement infinitely reducible first class
constraints up to a constraint system of finite stage of
reducibility has been proposed [19]. Before trying it within the
framework of BFV quantization, it seems natural to
consider canonical quantization of the model.

In the present letter we examine the problem in ${R}^{4/4}$
superspace and show that an effective field theory which corresponds to
the first quantized Siegel superparticle is the
massless Wess--Zumino model [20] in the component form.

\vspace{0.2cm}

2. The action functional which describes the dynamics of the Siegel
superparticle in ${R}^{4/4}$ superspace reads [21] (we use the spinor
notation from Ref. 22)
\begin{equation}
S=\int d\tau \{\frac 1{2e}\Pi^m\Pi_m +i\dot\theta\rho -i\bar\rho
\dot{\bar\theta}\},
\end{equation}
with
\[ \Pi^m=\dot x^m-i\theta\sigma^m\dot{\bar\theta}+i\dot\theta
\sigma^m\bar\theta+ i\psi\sigma^m\bar\rho-i\rho\sigma^m\bar\psi. \]
It is invariant under global supersymmetry transformations,
\begin{equation}
\delta\theta=\epsilon, \qquad \delta\bar\theta=\bar\epsilon, \qquad
\delta x^n=i\epsilon\sigma^n\bar\theta-i\theta\sigma^n\bar\epsilon,
\end{equation}
as well as under local $\alpha$- and $\kappa$-symmetries,
$$
\begin{array}{lll} \delta_\alpha\theta=\alpha\dot\theta, &
\delta_\alpha\bar\theta=\alpha\dot{\bar\theta}, &
\delta_\alpha x^n=\alpha\dot x^n,\\
\delta_\alpha\rho=\alpha\dot\rho, &
\delta_\alpha\bar\rho=\alpha\dot{\bar\rho}, &
\delta_\alpha e=(\alpha e)^\cdot,\\
\delta_\alpha\psi=(\alpha\psi)^\cdot, &
\delta_\alpha\bar\psi=(\alpha\bar\psi)^\cdot;\end{array}
\eqno{(3.a)}$$
$$
\begin{array}{l}
\delta_\kappa\theta=\displaystyle \frac 1e \Pi_n\sigma^n\bar\kappa,
\qquad \delta_\kappa\bar\theta=\frac 1e \Pi_n\kappa\sigma^n,\\
\delta_\kappa x^n=i\theta\sigma^n\delta\bar\theta-
i\delta\theta\sigma^n\bar\theta-i\kappa\sigma^n\bar\rho+
i\rho\sigma^n\bar\kappa,\\
\delta_\kappa e=4i\dot\theta\kappa-4i\bar\kappa\dot{\bar\theta}, \qquad
\delta_\kappa\psi=\dot\kappa,\\
\delta_\kappa\bar\psi=\dot{\bar\kappa}.\end{array}
\eqno{(3.b)}$$
\addtocounter{equation}{1}
As is seen from Eqs. (2) and (3), the coordinates $(x^m,\theta^\alpha,
\bar\theta_{\dot\alpha})$ parametrize the standard ${R}^{4/4}$
superspace, while the variables $e$ and $(\psi^\alpha,
\bar\psi_{\dot\alpha})$ prove to be the gauge fields for the local
$\alpha$- and $\kappa$-symmetries respectively. The role of the pair
$(\rho^\alpha, \bar\rho_{\dot\alpha})$ is to provide the terms
corresponding to (mixed) covariant propagator for fermions in the
action (1) [23].

Passing to the Hamiltonian formalism one finds primary constraints in
the problem
\addtocounter{equation}{1}
$$
\begin{array}{ll} p_\theta+ip_n\sigma^n\bar\theta+i\rho=0, & p_\rho=0,\\
p_{\bar\theta}+i\theta\sigma^np_n+i\bar\rho=0, & p_{\bar\rho}=0,
\end{array}
\eqno{(4.a)}$$
$$
p_{e}=0, \qquad p_{\psi}=0, \qquad p_{\bar\psi}=0,
\eqno{(4.b)}$$
where $(p,p_\theta,p_{\bar\theta},p_\psi,p_{\bar\psi},p_\rho,
p_{\bar\rho},p_e)$ are momenta conjugate to the variables \linebreak
$(x,\theta,\bar\theta, \psi,\bar\psi,\rho,\bar\rho,e)$ respectively.
The total Hamiltonian has the
form
\begin{eqnarray}
\lefteqn{H=p_e\lambda_e+p_\psi\lambda_\psi-\lambda_{\bar\psi}p_{\bar\psi}
+p_\rho\lambda_\rho-\lambda_{\bar\rho}p_{\bar\rho}+}\cr
&& +(p_{\bar\theta}+i\theta\sigma^np_n+i\bar\rho)\lambda_{\bar\theta}-
\lambda_\theta(p_\theta+ip_n\sigma^n\bar\theta+i\rho)+\cr
&& +e\displaystyle\frac{p^2}2-i\psi\sigma^np_n\bar\rho +
i\rho\sigma_np^n\bar\psi,
\end{eqnarray}
where $\lambda_{\dots}$ are Lagrange multipliers corresponding to the
primary constraints. The consistency conditions for the primary constraints
imply the secondary ones
\begin{equation}
p^2=0, \qquad p_n\sigma^n\bar\rho=0, \qquad \rho\sigma^np_n=0,
\end{equation} and determine some of the Lagrange multipliers,
\begin{equation}
\begin{array}{ll} \lambda_\theta=p_n\sigma^n\bar\psi,& \qquad
\lambda_{\bar\theta}=\psi\sigma^np_n,\\
\lambda_\rho=-2p_n\sigma^n\lambda_{\bar\theta}\approx0,& \qquad
\lambda_{\bar\rho}=-2\lambda_\theta\sigma^np_n\approx0.
\end{array}
\end{equation}
No tertiary constraints arise from the Dirac procedure and the remaining
Lagrange multipliers stay undetermined. Thus, the complete constraint
set of the model is given by Eqs. (4),(6) and it is convenient to rewrite
the latter in the equivalent form
\begin{equation}
p^2=0, \qquad p_m\tilde\sigma^mp_\theta=0, \qquad
p_{\bar\theta}\tilde\sigma^mp_m=0.
\end{equation}
The constraints (4.a) are second-class which means that the pairs
$(\rho,p_\rho)$, $(\bar\rho,p_{\bar\rho})$ can be omitted after
introducing the corresponding Dirac bracket\footnote{We define the
fundamental Poisson brackets as $\{x^m,p_n\}={\delta^m}_n$,
$\{\theta^\alpha,p_{\theta\beta}\}={\delta^\alpha}_\beta$,
$\{\bar\theta^{\dot\alpha},p_{\bar\theta\dot\beta}\}=
\delta^{\dot\alpha}{}_{\dot\beta}$.} (the Dirac brackets for the
remaining variables $\Gamma^i$ prove to coincide with the Poisson ones
$\{\Gamma^i,\Gamma^j\}_D$=$\{\Gamma^i,\Gamma^j\}$).
The first-class constraints (4.b) admit the covariant gauge
\begin{equation}
e =1, \qquad \psi=0, \qquad \bar\psi=0,
\end{equation}
which implies
\begin{equation}
\lambda_e=0, \qquad \lambda_\psi=0, \qquad \lambda_{\bar\psi}=0,
\end{equation}
and allows one to eliminate the variables $(e,p_e)$, $(\psi,p_\psi)$,
$(\bar\psi,p_{\bar\psi})$ from the consideration. Thus, after partial
phase space reduction, there remain $(x,p)$,
$(\theta,p_\theta)$, $(\bar\theta,p_{\bar\theta})$ variables in the problem
which are subject to the first-class constraints (8). The total Hamiltonian
vanishes on the constraint surface in full agreement with the
reparametrization invariance of the model.

Some remarks are relevant here. First, from Eq. (8) it follows
\begin{equation} p_\theta^2=0, \qquad
p_{\bar\theta}^2=0.
\end{equation}
This means that the $C$-constraint of the
$10d$ $ABCD$ superparticle [16] (which removes negative norm states from
the quantum spectrum [15-17]) automatically holds in four dimensions.
Second, the constraints (8) just
coincide with the first-class ones of the $4d$ Brink--Schwarz
superparticle [15]. It is infinitely reducible constraint set which
requires an infinite tower of ghost variables in BFV quantization [12,
15]. Third, a covariant gauge to Eq. (8) is known to be problematic in the
original phase space [10] (the conventional noncovariant gauge choice is
$x^+=\tau p^+$, $\theta\sigma^+=0$, $\sigma^+\bar\theta=0$). We refrain
from fixing the gauge freedom related to the first-class constraints (8)
and will treat them as restrictions on physical states in the course of
Dirac quantization. This gives us a possibility to develop a covariant
scheme of quantization which is presented below.

\vspace{0.2cm}

3. Since Dirac brackets for the independent variables coincide with the
Poisson ones, the commutation relations for the corresponding operators are
canonical. We can realize them in the coordinate representation,
\begin{eqnarray}
&& \hat x^n=x^n , \qquad \hat p_m=-i\displaystyle\frac\partial{\partial x^m},
\qquad [\hat x^n,\hat p_m]=i{\delta^n}_m,\cr
&& \hat\theta^\alpha=\theta^\alpha, \qquad \hat p_{\theta_\alpha}=
i\displaystyle\frac\partial{\partial\theta^\alpha}, \qquad
[\hat\theta^\alpha,\hat p_{\theta_\beta}]=i{\delta^\alpha}_\beta,\\
&& \hat{{\bar\theta}^{\dot\alpha}}={\bar\theta}^{\dot\alpha} ,
\qquad \hat p_{\bar\theta_{\dot\alpha}}
=i\displaystyle\frac\partial{\partial\bar\theta^{\dot\alpha}}, \qquad
[\hat{{\bar\theta}^{\dot\alpha}},\hat p_{\bar\theta_{\dot\beta}}]=
i\delta^{\dot\alpha}{}_{\dot\beta}.\nonumber
\end{eqnarray}
on a Hilbert space whose elements are chosen to be real scalar superfields
(they do not depend on time since the Hamiltonian vanishes on the
constraint surface)
\begin{eqnarray}
\lefteqn{V(x,\theta,\bar\theta)=A(x)+\theta\psi(x)+
\bar\theta\bar\psi(x)+\theta^2F(x)+\bar\theta^2
\bar F(x)+}\cr
&& +\theta\sigma^n\bar\theta C_n(x)+\bar\theta^2\theta\lambda(x)
+\theta^2\bar\theta\bar\lambda(x)+
\theta^2\bar\theta^2D(x),
\end{eqnarray}
and which obey the boundary condition
\begin{equation}
V(x,\theta,\bar\theta)\mathop{\longrightarrow}\limits_{x\to\pm\infty}0\;.
\end{equation}

The physical states in the Hilbert space are defined in the usual way
[24,25]
\begin{eqnarray} && \hat p^2|{\rm phys}\rangle=0,\cr &&
\tilde\sigma^n\hat p_n\hat p_\theta|{\rm phys}\rangle=0,\\ &&
\tilde\sigma^n\hat p_n\hat p_{\bar\theta}|{\rm phys}\rangle=0.\nonumber
\end{eqnarray}
In the explicit representation (12), (13) this yields
$$
\begin{array}{l} \tilde\sigma^n\partial_n\psi=0, \qquad
\tilde\sigma^n\partial_n\bar\psi=0,\\
\Box A=0;\end{array}
\eqno{(16.a)}$$
$$
\begin{array}{l}
(\tilde\sigma^m\sigma^n)^{\dot\alpha}{}_{\dot\beta}\partial_m C_n=0,
\qquad (\sigma^n\tilde\sigma^m)_\alpha{}^\beta\partial_m C_n=0,\\
\Box C_n=0,
\end{array}
\eqno{(16.b)}$$
\addtocounter{equation}{1}
with all other component fields vanishing due to the boundary condition
(14). In obtaining Eq. (16) the identity
\begin{equation}
{\rm Tr}\,(\sigma^n\tilde\sigma^m)=-2\eta^{nm}
\end{equation}
was used.

Consider now Eq.(16.b). Taking a trace of the first equation and
making use of Eq. (17) one finds
\begin{equation}
\partial^nC_n=0,
\end{equation}
which (with the use of the standard relation $\sigma^n\tilde\sigma^m+
\sigma^m\tilde\sigma^n=-2\eta^{nm}$) allows one to rewrite Eq. (16.b)
in the form
\begin{equation}
\begin{array}{l}
(\sigma^{mn})_\alpha{}^\beta\partial_m C_n=0, \qquad
(\tilde\sigma^{mn})^{\dot\alpha}{}_{\dot\beta}\partial_m C_n=0,\\
\Box C_n=0\end{array}
\end{equation}
Multiplying the first equality in Eq. (19) with
$(\sigma^{kl})_\beta{}^\alpha$, and taking into account the identity
\begin{equation}
{\rm Tr}\,\sigma^{mn}\sigma^{kl}=-\frac 12(\eta^{mk}\eta^{nl}-
\eta^{ml}\eta^{nk})-\frac i2\epsilon^{mnkl},
\end{equation}
one gets
\begin{equation}
\partial_kC_l-\partial_lC_k=-i\epsilon_{klmn}\partial^mC^n,
\end{equation}
which together with its complex conjugate implies
\begin{equation}
\partial_mC_n-\partial_nC_m=0, \qquad \epsilon_{klmn}\partial^mC^n=0.
\end{equation}
The only solution to Eqs. (18), (22) is
\begin{equation}
C_n=\partial_nB,
\end{equation}
with $B$ the on-shell massless scalar field
\begin{equation}
\Box B=0.
\end{equation}
Thus, physical states of the first quantized Siegel superparticle look
like
\begin{equation} V_{\rm phys}(x,\theta,\bar\theta)=A(x)+\theta\psi(x)
+\bar\theta\bar\psi(x)+\theta\sigma^n\bar\theta
\partial_nB(x),
\end{equation}
with $A,B$ the on-shell massless real scalar fields (irreps of the
Poincar\'e group of helicity 0) and $\psi,\bar\psi$ the on-shell massless
spinor fields (helicities 1/2 and --1/2, respectively). Note that together
they fit to form two irreducible representations of the super Poincar\'e
group of superhelicities 0 and --1/2 [26].

It is worth mentioning that Eq. (15) can be rewritten in the manifestly
superinvariant form
\begin{equation}
\tilde\sigma^{n\dot\alpha\alpha}\partial_nD_\alpha V=0, \qquad
\tilde\sigma^{n\dot\alpha\alpha}\partial_n\bar D_{\dot\alpha}V=0,
\end{equation}
where $D_\alpha$, $\bar D_{\dot\alpha}$ are the covariant derivatives,
or as the single massless Dirac equation
\begin{equation}
\gamma^n\partial_n\Psi=0,
\end{equation}
with $\Psi\equiv\left(\begin{array}{c} D_\alpha V\\
\bar D^{\dot\alpha}V\end{array}\right)$ a Majorana (superfield) spinor.

An effective field theory which reproduces equations (16.a), (24) is
easy to write
\begin{equation}
S=\int d^4x\Big\{\frac 12 \partial^mA\partial_mA+ \frac 12 \partial^mB
\partial_mB+i\psi\sigma^m\partial_m\bar\psi\Big\},
\end{equation}
which is invariant under the global (on-shell) supersymmetry
transformations
\begin{eqnarray}
&& \delta A=\epsilon\psi+\bar\epsilon\bar\psi, \qquad
\delta B=i\epsilon\psi-i\bar\epsilon\bar\psi,\cr
&& \delta\psi=i(\sigma^n\bar\epsilon)\partial_nA+
(\sigma^n\bar\epsilon)\partial_nB,\\
&& \delta\bar\psi=-i(\epsilon\sigma^n)\partial_nA+
(\epsilon\sigma^n)\partial_nB.\nonumber
\end{eqnarray}
In Eq. (28) we recognize the massless Wess--Zumino model in the component
form [20].

Thus, we have demonstrated that first quantization of the Siegel
superparticle leads to the massless Wess-Zumino theory.

\vspace{0.2cm}

4. As is known, the superfield formulation of the massless Wess--Zumino
model involves chiral and antichiral superfields [20, 22],
$$ S = \int d^8z\, \Phi\bar\Phi,
\eqno{(30.a)}$$
$$ \bar D_{\dot\alpha}\Phi=0,
\eqno{(30.b)}$$
$$ D_\alpha\bar\Phi=0.
\eqno{(30.c)}$$
Equations of motion for the theory (30) read
$$ D^2\Phi=0,
\eqno{(31.a)}$$
$$ \bar D^2\bar\Phi=0.
\eqno{(31.b)}$$
\addtocounter{equation}{2}
Let us  show that the real scalar superfield (13) satisfying the
constraints (26) is the sum of on-shell massless chiral ((30.b), (31.a))
and antichiral ((30.c), (31.b)) superfields.

Consider Eqs. (30.b), (31.a). The first of them implies the decomposition
[22, 26]
\begin{equation}
\Phi(x,\theta,\bar\theta)=\alpha(x)+\theta\psi(x)+\theta^2f(x)+
i\theta\sigma^n\bar\theta\partial_n\alpha(x)+\frac 12 \theta^2
\bar\theta\tilde\sigma^n\partial_n\psi+\frac 14\theta^2
\bar\theta^2\Box\alpha,
\end{equation}
while the latter, being rewritten in the equivalent form
\begin{equation}
\tilde\sigma^{n\,\dot\alpha\alpha}\partial_nD_\alpha\Phi=0,
\end{equation}
yields
\begin{equation}
\tilde\sigma^n\partial_n\psi=0, \qquad \Box\alpha=0, \qquad f=0.
\end{equation}
To get Eqs. (33), (34) we used the identity
\begin{equation}
[D^2,\bar D_{\dot\alpha}]=-4i{\sigma^n}_{\alpha\dot\alpha}
\partial_nD^\alpha,
\end{equation}
and assumed the standard boundary conditions. Note also that the
chirality condition together with Eq. (33) implies
\begin{equation}
\Box\Phi=0,
\end{equation}
as a consequence of
\begin{equation}
\{ D_\alpha,\bar D_{\dot\alpha}\}=-2i\sigma^n{}_{\alpha\dot\alpha}
\partial_n.
\end{equation}
Thus, an on-shell scalar chiral superfield can be written as
\begin{equation}
\begin{array}{c}
\Phi(x,\theta,\bar\theta)=\alpha(x)+\theta\psi(x)+i\theta\sigma^n\bar\theta
\partial_n\alpha(x),\\
\Box\alpha(x)=0, \qquad \tilde\sigma^n\partial_n\psi(x)=0.\end{array}
\end{equation}
Similarly, an on-shell scalar antichiral superfield ((30.c), (31.b)) reads
\begin{equation}
\begin{array}{c}
\bar\Phi(x,\theta,\bar\theta)=\bar\alpha(x)+\bar\theta\bar\psi(x)-
i\theta\sigma^n\bar\theta\partial_n\bar\alpha(x),\\
\Box\bar\alpha(x)=0, \qquad \tilde\sigma^n\partial_n\bar\psi(x)=0.\end{array}
\end{equation}
Considering now the sum
\begin{equation}
\Phi+\bar\Phi=(\alpha+\bar\alpha)+\theta\psi+\bar\theta\bar\psi+
\theta\sigma^n\bar\theta\partial_ni(\alpha-\bar\alpha),
\end{equation}
and denoting
\begin{equation}
\alpha+\bar\alpha=A, \qquad i(\alpha-\bar\alpha)=B,
\end{equation}
one arrives just at Eq. (25). Thus, the real scalar superfield
subject to the constraints (26) was proven to be the sum of on-shell chiral
and antichiral superfields
\begin{equation} V_{\rm
phys}(x,\theta,\bar\theta)=\Phi(x,\theta,\bar\theta)+
\bar\Phi(x,\theta,\bar\theta).
\end{equation}
As is known, on-shell massless scalar chiral superfields form massless
irreducible representation of the super Poincar\'e group of
superhelicity $0$ [26]. Analogously, on-shell massless scalar antichiral
superfields realize irrep of superhelicity $-1/2$. We may conclude here
that quantum states of the first quantized
Siegel superparticle form a reducible representation of the super
Poincar\'e group which contains superhelicities 0 and $-1/2$.

\vspace{0.4cm}
Thus, in this letter we have considered canonical quantization of the
Siegel superparticle in ${R}^{4/4}$ superspace.. Quantum states of the
model were proven to be the sum of on-shell chiral and antichiral
superfields.  The corresponding effective field theory was shown to be the
massless Wess--Zumino model. Because propagators of the theory are well
known, it seems interesting to reproduce Siegel's action within the
framework of the proper-time approach [27], as well as, to compare
the result with that of the straightforward BFV quantization combined
with the scheme [19]. This work is in progress now.

As was mentioned above, the $C$-constraint of the $10d$
$ABCD$-su\-per\-particle [16] is not necessary in four dimensions. It is
tempting to extend the model (1) up to a theory equivalent to the $4d$
Brink--Schwarz superparticle along the lines of Ref. 16.
Since constraints quadratic in fermions will no appear now, we expect
considerable simplification in the corresponding BFV quantization.

Due to the relation to superstring theory, the $10d$ case is of prime
interest. The operatorial quantization presented in this work is rather
specific in four dimensions. We hope, however, that BFV path integral
quantization will proceed along the same lines both in $4d$ and $10d$.
The results on this subject will be present elsewhere.

\section*{Acknowledgments}
A.A.D. thanks the Institute for Theoretical Physics of Hannover University
for hospitality.The work of A.A.D has been supported in part by the
Joint DFG--RFBR project No 96--02--00180G. A.V.G. thanks the ICTP for
the hospitality at the early stage of this work. D.M.G. thanks Brasilian
foundation CNPq for permanent support.

\end{document}